# Photoinduced split of the cavity mode in photonic crystals based on porous silicon filled with photochromic azobenzene-containing substances

Alexey Bobrovsky[1*], Sergey Svyakhovskiy[2], Valery Shibaev[1], Martin Cigl[3], Vera Hamplová[3], Alexej Bubnov[3]

[1]*Department of Chemistry, M.V. Lomonosov Moscow State University, 1 Leninskie gory, Moscow, 119991, Russia.*

[2]*Department of Physics, M.V. Lomonosov Moscow State University, 1 Leninskie gory, Moscow, 119991, Russia,*

[3]*Institute of Physics of the Czech Academy of Sciences, 1999/2 Na Slovance, 182 21 Prague 8, Czech Republic*

The novel phototunable photonic structures based on electrochemically etched silicon filled with four photochromic azobenzene-containing compounds, bent-shaped low molar mass substance and side-chain polymethacrylates and copolyacrylate, were prepared and their photooptical properties were studied. It was found that irradiation of these composites with polarized blue light results in spectral changes in photonic band gap (split of the cavity mode) associated with cooperative photoorientation of azobenzene moieties inside silicon pores in direction perpendicular to the polarization plane of the incident light. Kinetics of the photoinduced split is studied. The observed phototoinduced split is completely reversible and heating of the composites to temperatures above isotropization or glass transitions fully recovers the initial spectral shape of photonic band gap. Thermal and temporal stability of the obtained photoinduced split were comparatively studied, and it was found that for composites with bent-shaped substance and polymethacrylate shape of the reflectance spectra does not change over time at room temperature. The prepared composites have high potential for the different applications in photonics.

**Introduction**

Design and investigation of stimuli-responsive photonic structures are promising and fast developed topical areas in the modern science and technology. Photonic crystals (PhCs) have gained much interest in recent years due to a large variety of possible applications in photonics and optoelectronics and other related fields of technology [1]. Among different type of these structures, photocontrollable photonic crystals (PhCs) are of a special interest because the light is very a convenient and contactless tool for the manipulation of structure and optical properties.

One of the efficient ways of implementation PhCs phototunability implies infiltration of inverse opal structures or porous PhCs with photochromic liquid crystalline (LC) mixtures. There are several papers describing photosensitive PhCs, in which the phototunability is provided by photoinduced isothermal phase transition [2-5]. In most cases, UV light induced E-Z isomerization leads to the formation of bent-shaped Z isomers which destabilize LC phase and induce isothermal LC phase – isotropic state transition. As a result, a noticeable change in the refractive index of photochromic mixtures is observed, with a shift of the photonic band gap (PBG) position or changing of the PhC reflectivity. As example, such hybrid PhCs were obtained using stretched poly(methyl methacrylate) (PMMA) inverse opal filled with nematic LC mixture of 4-pentyl-4'-cyanobiphenyl (5CB) and photochromic dopant 4-butyl-4'-methoxyazobenzene.[4] UV irradiation causes a blue shift of PBG position due to the change of the refractive index after mixture isotropization.

In another paper [5] photochromic LC mixture of nematic (E7) doped with 4-butyl-4′-methoxyazobenzene was introduced in one-dimensional PhC based on anodic aluminum oxide. It was shown, that UV irradiation induces red shift of the photonic band gap, and a good fatigue resistance of composite to ON-OFF cycles was demonstrated. It is noteworthy, that these composites suffer from significant drawbacks restricting their applicability, namely the poor quality of PBG structure, low reflectance intensity and fast thermal Z-E isomerization of selected azobenzene dopant leading to a quick recovery of initial PhC spectra. Recovery time was found as 1.44 s at the temperature of 35 °C. This casts doubt on the mechanism of photo-induced changes proposed by the authors because such short lifetime of Z-form is impossible for the azobenzene substance of this type.

In our recent paper [6] we have demonstrated an efficient photochemically reversible phototuning of photonic band gap in PhC having highly defined photonic structure with cavity mode. PhCs obtained by electrochemical etching of crystalline silicon were filled with nematic mixtures of 4-pentyl-4'-cyanobiphenyl (5CB) doped with azobenzene-containing photochromic dopant with lateral methyl substituents providing a good solubility and a high thermal stability of photoinduced Z-isomer. A noticeable reversible shift of photonic band gap and cavity mode position (ca. 10 nm) is shown under UV-light irradiation.

Another efficient way of phototuning the structure and optical properties of photochromic glass forming substances is so-called Weigert effect representing photoorientation of the azobenzene chromophores under polarized light action [7-11]. Photoinduced repetitive cycles of E-Z-E isomerization accompanied by rotational diffusion of chromophores align the chromophores in the direction orthogonal to the polarization of incident excitation light (**Figure 1**). In amorphous polymers or glass-forming low-molar-mass substances this effect enables to obtain high values of dichroism or birefringence [7-11]. In polydomain LC samples, uniaxial alignment of mesogenic groups could be achieved (although ordered LC state in some cases prevents photoorientation process, especially, for the smectic phase [11]).

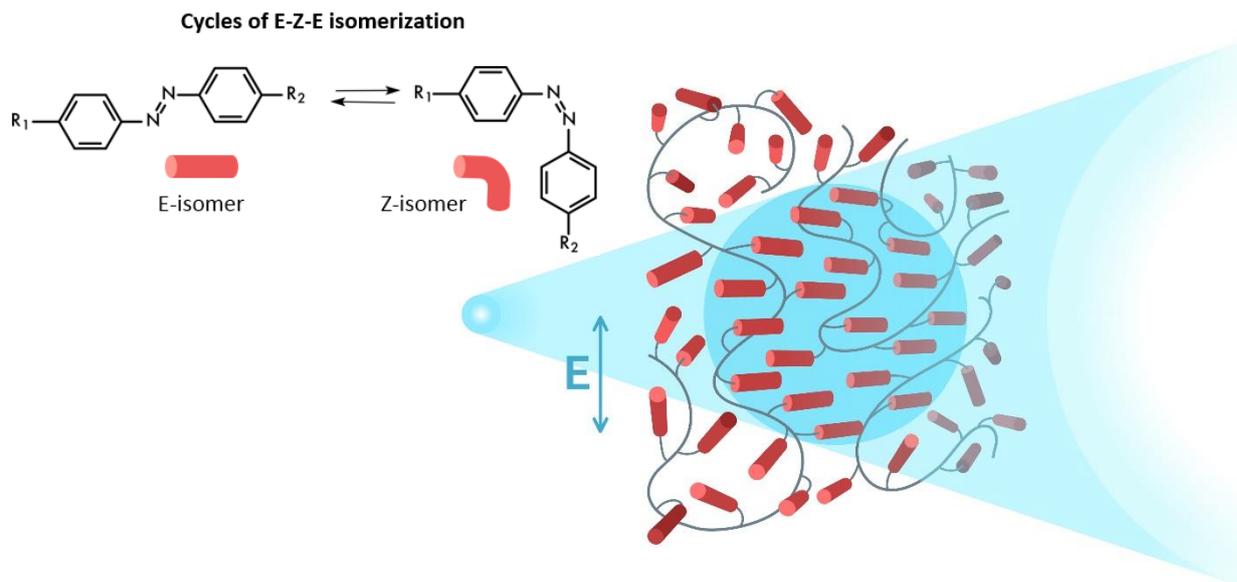

**Figure 1.** Illustration of the photoorientation process of the chromophores induced by polarized light action.

In the present paper, we have developed an approach of the reversible PBG phototuning based on photoorientation of azobenzene chromophores under polarized visible light action. As it was shown photoorientation process results in an efficient cavity mode split due to the appearance of the birefringence. As a template for PhC preparation electrochemically etched silicon was selected, the similar as was used in our previous work [6]. Scheme of the PhC structure is shown in **Figure 2a.**

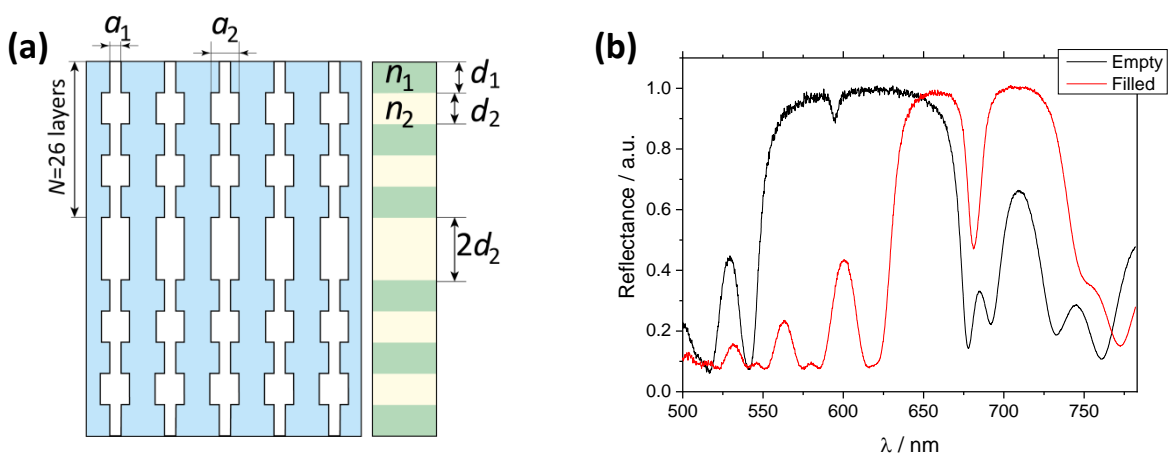

**Figure 2.** (a) The scheme of the PhC structure and its geometric characteristics. The diameters of pores are $a_1=20\pm5$ nm and $a_2=60\pm5$ nm, the thicknesses of layers are $d_1=96$ nm and $d_2=108$ nm, the cavity layer thickness $d_C=2d_2=216$ nm. (b) Reflectance spectra of the porous Si photonic structure before and after filling with bent-shaped mesogenic compound **1**.

In the prepared silicon matrix the pores obtained by electroetching are aligned exactly along normal to the silicon plate. Their diameters measured by the low-temperature nitrogen desorption method (see Experimental part) are about $a_1$=20 nm and $a_2$=60 nm, respectively. The layers of low porosity ($a_1$) have the high refractive index of $n_1$=2.2, whereas the layers of higher porosity ($a_2$) have the low refractive index of $n_2$=1.8. PhC consists of also microcavity layer of the thickness of 216 nm located between two Bragg mirrors of 18 layers each. Reflectance spectra of the porous PhC is presented in Figure 2b demonstrating well-defined reflectance peak located in visible spectral range and cavity mode at 594.6 nm.

This porous silicon PhC was filled with different low-molar-mass and macromolecular substances. Structures of these compounds are shown in **Figure 3**; the phase behavior and molecular mass characteristics are presented in **Table 1**. Substance **1** is glass-forming azobenzene-containing molecules with bent-shaped molecular geometry. Its photooptical properties were studied in our previous papers [12, 13] showing excellent photoinduced dichroism in amorphousized films obtained by spin-coating. As polymeric substances two azobenzene-containing homopolymers **P1** and **P2** and copolymer **P3** were synthesized. Their photoorientation behavior under UV and visible polarized light was also well-studied by us before [12, 14-16].

**Table 1.** Phase behavior and molecular masses of the azobenzene-containing substances (temperatures in ºC are presented). Notations: I – stands for the isotropic liquid; N – is the nematic phase; N* - in the chiral nematic (cholesteric) phase; SmC* - is the tilted ferroelectric smectic phase. The phase transition temperatures are presented on heating.

| Material | Phase transitions | $M_w$ | $M_w/M_n$ |
|---|---|---|---|
| **1** | Cr 181 N 200 I | - | - |
| **P1** | g 58 N 159 I | 11600 | 1.6 |
| **P2** | SmC* 176 N* 228 I | 15400 | 1.9 |
| **P3** | g 28 N 123 I | 8300 | 1.3 |

The main objective of this work is to demonstrate of the possibility of phototuning and photoinduced split of the cavity mode under polarized light action. The influence of different factors, such as the structure of the selected azobenzene-containing substances, their phase behavior and wavelength of the excitation light was studied. The thermal stability of the photoinduced changes in reflection spectra was examined and compared for the different samples.

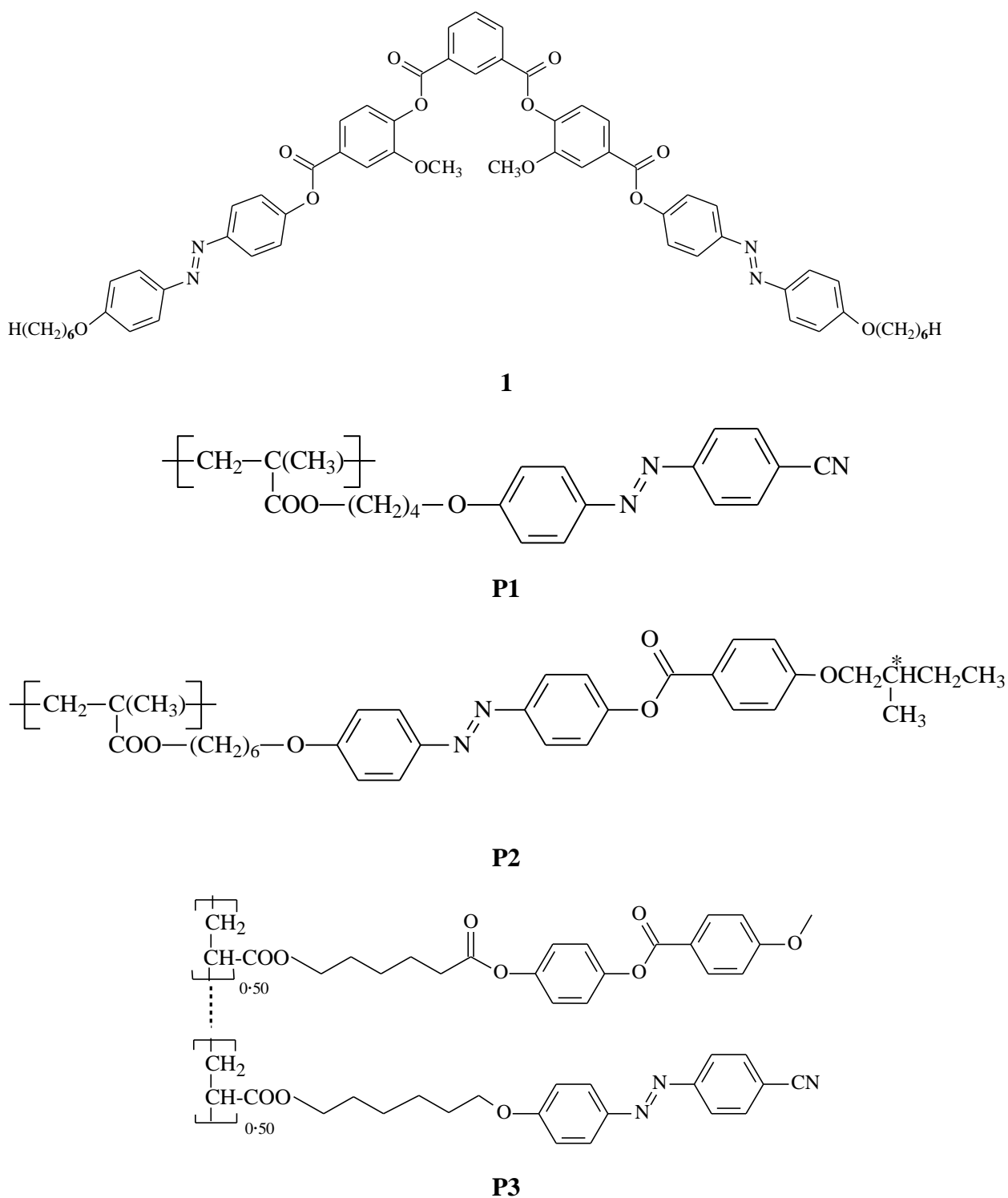

**Figure 3.** Structures of photochromic azobenene-containing substances under study.

**Results and discussion**

Filling of porous Si with photochromic substances shifts PBG to longer wavelength range (**Figures 2b, S1**, values of $\lambda_0$ in **Table 2**) because air inside pores is replaced with compounds having higher refractive index ($<n> \sim 1.6$-$1.7$). An amplitude of the shift is maximal for the copolymer **P3** and predetermined by several factors including refractive indices of the copolymer, alignment of the chromophores inside pores and degree of filling.

Irradiation of all composite samples with polarized blue light (457 nm) results in split of the cavity mode of the PBG (**Figures 4**; values of Δλ in **Table 2**). Such duplication of the cavity mode is explained by different refractive indices for different polarization directions. As seen from **Figures 5**, the position of the cavity mode measured for reflected light polarized perpendicular to the plane of the polarization of the laser is shifted to the longer wavelengths in respect to the parallel one. In other words, the refractive index for light polarized in a direction perpendicular to the polarization of light used for photoorientation is higher. This is the consequence of the photoorientation process of the chromophores in direction perpendicular to the light polarization (see schemes in **Figures 1, 4d**).

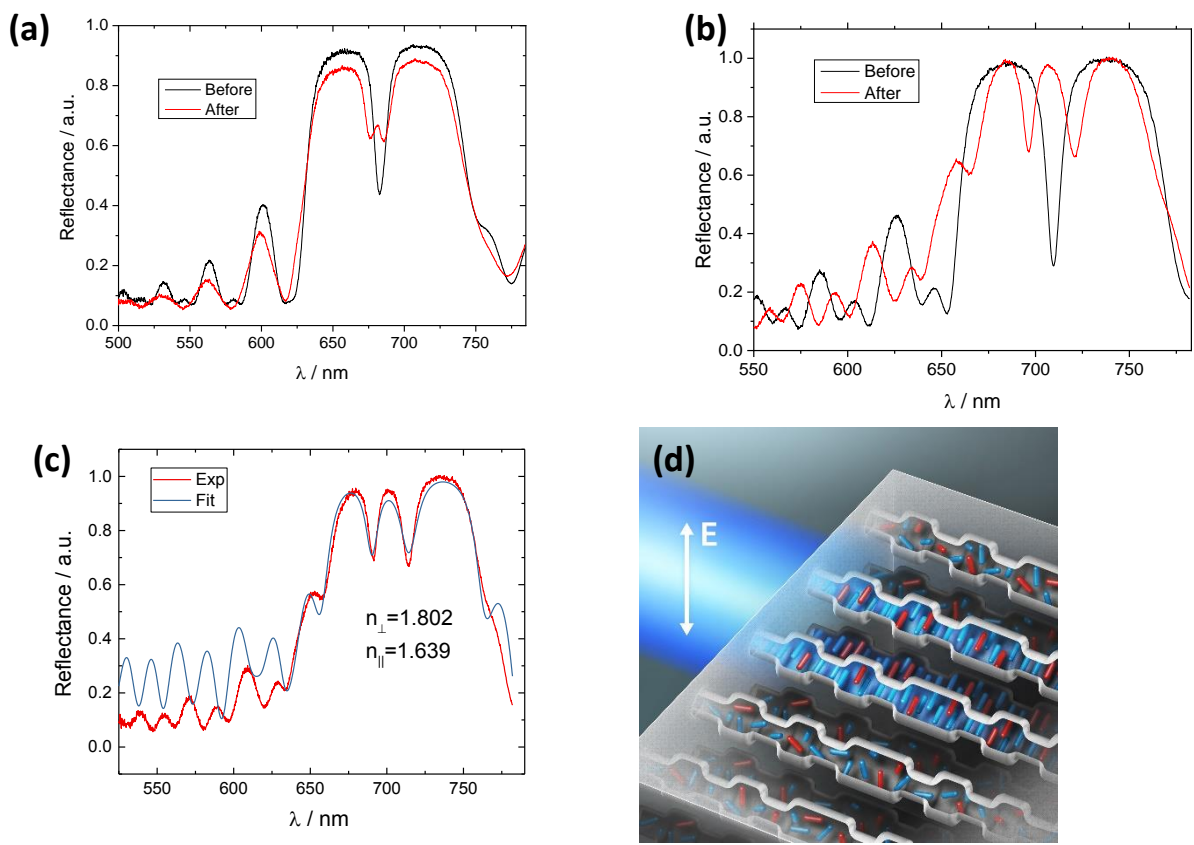

**Figure 4.** Nonpolarized reflectance spectra of composites with **1** (a) and **P3** (b) before and after irradiation (80 min, 457 nm, ~300 mW/cm$^2$). (c) The theoretical and experimental spectra of the irradiated **P3** sample. The parameters of the theoretical approximation were the refractive indices in the parallel ($n_\parallel$) and perpendicular ($n_\perp$) directions (written in figures). (d) Scheme of the alignment of chromophores (red) and mesogenic groups (blue) inside pores of PhC; arrow shows polarization direction.

The reflectance spectrum for the irradiated samples of the composite with copolymer **P3** was fitted with the theoretical model (see Experimental part for the details of calculation). **Figure 4c** shows the experimental spectrum of the irradiated **P3** sample and its fit with the theoretical model. The model includes a biaxially anisotropic composite medium as the filling material of the pores of the PhC. For the composite sample the refractive indices $n_\perp$=**1.802** and $n_\parallel$=**1.639**, birefringence Δn=**0.163**.

**Table 2.** Position of the cavity mode λ₀ before irradiation and values of the maximal cavity mode split (Δλ) after irradiation with 457 nm laser. Experimental error in wavelength determination does not exceed ±0.5 nm.

| Composite | $\lambda_0$ (nm) | $\Delta\lambda$ (nm) |
|---|---|---|
| **1** | 681.4 | 10.4 |
| **P1** | 695.2 | 3.1 |
| **P2** | 667.4 | 2.6 |
| **P3** | 709.6 | 24.8 |

At the same time, refractive indices and birefringence of the photochromic copolymer **P3** were determined by polarized absorbance measurements of uniaxially aligned polyimide-coated electrooptical cell: extraordinary $n_e$=**1.903±0.033**, ordinary $n_o$=**1.570±0.022**; birefringence $\Delta n = n_e - n_o$=**0.333±0.055**. Comparison of these results shows that photoinduced alignment of the chromophores in porous composites is about twice as less in comparison with rubbed glass cell. This can be explained by the critical influence of the confinements of the chromophores inside the pores preventing the photoorientation.

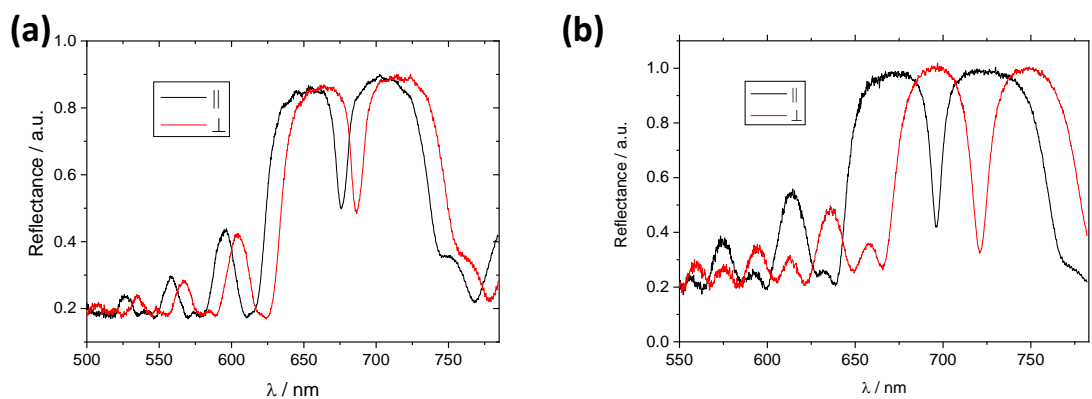

**Figure 5.** Polarized reflectance spectra of composites with **1** (a) and **P3** (b) after irradiation (457 nm). Irradiation time is 80 min. Reflectance was measured in directions parallel and perpendicular to the polarization of the laser light.

We have studied kinetics curves of changes of the cavity mode position for different polarizations of reflected light (**Figure 6a**) and growth of cavity mode split (**Figure 6b**) during irradiation. As seen from these figures photostationary state is achieved after ca. 60 min of the irradiation. Maximal values of the cavity mode split Δλ for all samples are presented in Table 2.

Highest Δλ was obtained for composite based on the copolymer **P3**, ca. 25 nm. Most probably, this is explained by highest mobility of the azobenzene chromophores which undergo photoorientation process cooperatively with nematogenic phenylbenzoate groups of the copolymer [9]. For another samples higher concentration of the azobenzene chromophores leading to stronger absorbance could decrease efficiency of the photoorientation process. The lowest values obtained for the composite with smectic polymer **P2** is explained by an influence of ordered smectic phase strongly preventing rotational diffusion of the chromophores. Similar effect was observed repeatedly in previous studies [11] showing crucial effect of smectic phase formation on photoorientation.

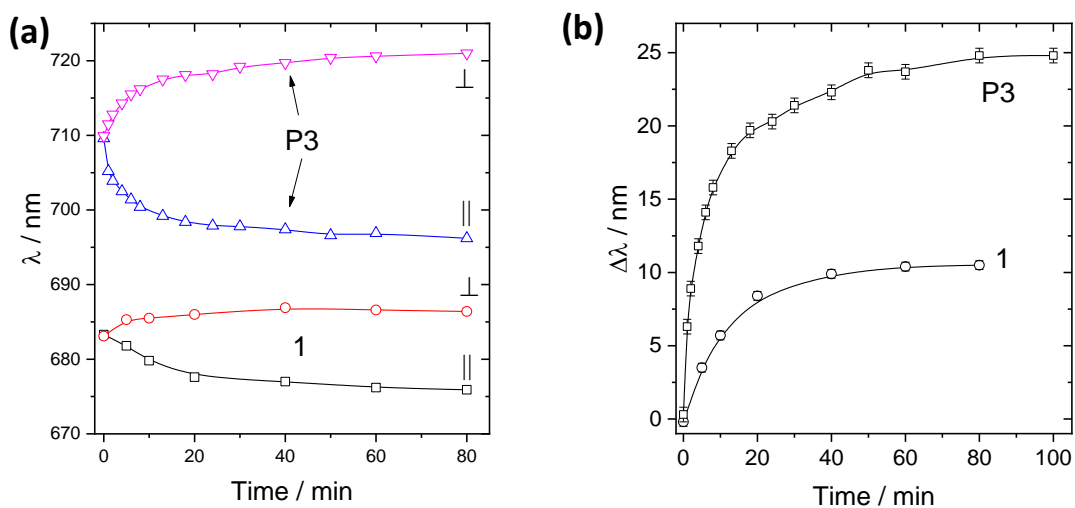

**Figure 6.** Kinetics of changes of the cavity mode position during irradiation (457 nm) for different polarizations of reflected light (a) and values of cavity mode split for of composite with **1** and **P3** (b).

A very interesting observation is related to high $\Delta\lambda$ values for composite with bent-shaped substance **1** (10.4). This effect is quite surprising because as was shown by us previously, for the crystalline films of this compound photoorientation process is completely suppressed by crystallization [12]. This leads us to the conclusion that crystallization process inside silicon pores does not take place at all or highly suppressed. However, this is not surprising because confinement often suppresses formation of crystallization nuclei [17] preventing crystallization process.

Irradiation with polarized UV light does not lead to the cavity mode split; long time irradiation induces photochemical degradation of the azobenzene-containing compounds and shift of the photonic band gap and cavity mode position to shorter wavelength range, closer to spectra of empty porous silicon PhC (**Figure S2**).

A stability of the photoinduced cavity mode split in time is studied. For this purpose samples were kept at room temperature and $\Delta\lambda$ values have been monitored. As seen from **Figures 7a**, composite with polymer **P1** demonstrates good thermal stability at room temperature, whereas for copolymer **P3** $\Delta\lambda$ slightly decreases even at room temperature. During first three days ca. 15% drop of $\Delta\lambda$ is observed followed by its stabilization (**Figure 7a**). Glass transition of the polymer **P3** is close to room temperature (28 ºC) that enables slow realignment of the chromophores toward to the initial orientation even at ambient conditions.

Composite with polymer **P1** demonstrates a good thermal stability of photoinduced $\Delta\lambda$ values because has higher glass transition temperature (58 ºC) (**Figure 7a**).

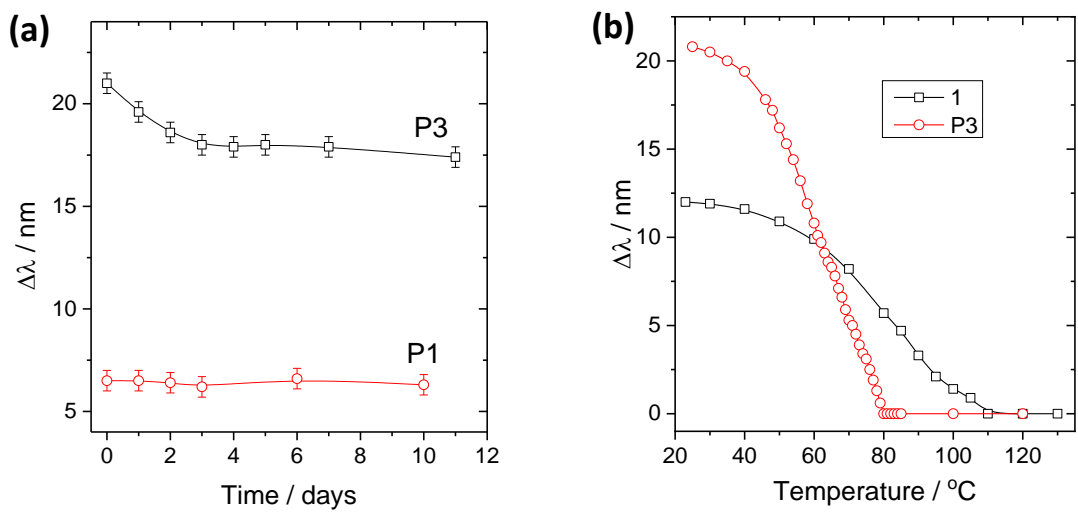

**Figure 7.** (a) Changes of the photoinduced cavity mode split at room temperatures for composites with **P1** and **P3**. (b) Dependence of the cavity mode split values on temperature for the composites with bent-shaped substance **1** and copolymer **P3**; heating rate 1 °/min.

Such relative instability of photoinduced Δλ for **P3** is quite unusual and unexpected, because as was shown before photoinduced alignment of photochromic and mesogenic groups in thin films is stable at room temperature [16]. Moreover, an annealing of the irradiated films at temperature higher the glass transition (50 °C) results in improvement of uniaxial orientation and increase in photoinduced dichroism, so-called "gain-effect". Contrary to thin films, under continuous heating of the irradiated composite samples with **P3** Δλ drops to zero at 80 °C (**Figure 7b**). Probably, such thermal instability of the photoinduced orientation is associated with effect of confinement or some sort of the "memory effect" which was described for the smectic azobenzene-containing polymers in one of our previous works [14]. Similarly, the complex pores topology contributes to fixing the initial alignment of chromophores and mesogens near the silicon-polymer interface.

Under heating, the photoinduced Δλ of composite containing photosensitive substance **1** decreases to zero at ca. 110 °C (**Figure 7b**). On the other hand Δλ the values for this composite are stable at room temperature for the very long time, i.e. examination of the irradiated sample after two years show insignificant changes.

**Conclusion**

Novel photocontrollable PhCs based on porous silicon matrix filled with photochromic azobenzene-containing glass-forming low-molar-mass substance and three polymers were obtained and their unique photooptical properties were demonstrated. The most remarkable effect of the polarized light action on such composite is the split of the PhC cavity mode due to the photoorientation process of the azobenzene groups in the direction orthogonal to the polarization plane of the excitation light (**Figures 1, 4d**). The largest values of split, ~25 nm was found for the copolymer containing azobenzene and phenylbenzoate side groups. The observed phenomenon was shown to be completely reversible and heating of the samples to temperatures above glass transition recovers initial PBG shape and position.

Our results clearly reveal how the introduction of photosensitive glass-forming substances with photochromic azobenzene moieties provides the ability to photocontrol the PBG shape and cavity mode position, which opens the door to the creation of advanced materials targeted for photonics and optoelectronics.

**Experimental section**

*Synthesis of photochromic substances and their characterization*

Polymers **P1, P2, P3** were synthesized by radical polymerization of the corresponding monomers according to [14-16] respectively. Synthesis of **6WAVI** is described in [12].

The phase transition temperatures of the monomers and polymers were detected by differential scanning calorimetry (DSC) using a Perkin Elmer DSC 8500. Dry nitrogen was purged through the DSC cell. The standard aluminium pans with 10 mg sample were used.

The polarizing optical microscope (POM) investigations were performed using LOMO P-112 polarizing microscope equipped by Mettler TA-400 heating stage. UV cut-off filter was used for preventing E-Z isomerization of azobenzene chromophores during POM observations.

*Porous silicon PhCs preparation*

1D photonic crystals were prepared by the porous silicon technique [18] using electrochemical etching of crystalline silicon wafers. Single-crystal wafers of the (100) surface orientation and a resistance of 5 mΩ×cm were etched in a two-electrode cell using a 28 wt% solution of hydrofluoric acid in ethanol as an electrolyte. The structure of the photonic crystal was formed by a periodic modulation of the current density by square wave function with levels of 40 mA/cm$^2$ and 200 mA/cm$^2$. This periodic modulation of the etching current density causes the formation of periodically alternating layers of different porosities. Corresponding diameters of pores are of about $a_1$=20 and $a_2$=60 nm which has been measured by low-temperature nitrogen adsorption-desorption isotherms (ASAP 2000 porometer, Micromeritics, United States). The measured refractive indices measured by thin-film interference were of $n_1$=2.2 and of $n_2$=1.8, respectively. The scheme of the sample cross-section is shown in **Figure 2a**. The microcavity layer of the thickness of 216 nm is located between two Bragg mirrors of 18 layers each.

*Theoretical calculations of the optical reflectance spectra*

The theoretical calculation of the optical spectra of the photonic crystal structure was performed by the optimized transfer matrix method[18] implemented in C code. Dielectric functions and refractive indices of the porous silicon layers were calculated by the effective medium approximation with Bruggeman model[19]. The model of the effective medium was modified by taking into account three components: silicon, liquid crystal with given dielectric constant, and the air. The refractive index of silicon was used from Ref. [20].

*Composite preparation*

For composite preparation drop of concentrated solution of substances in chloroform (~100 mg/mL) was placed on top of porous Si followed by slow evaporation. After drying at room temperature for several days. The excess of photosensitive substance was removed gently by paper at temperature ca. 20º higher isotropization transition followed by slow cooling of the obtained composite to room temperature (1 º/min).

*Photooptical investigations*

Irradiation was performed using MBL-N-457 diode laser (457 nm, ~300 mW/cm$^2$, CNI Laser) and AO-V-355 (355 nm, ~100 mW/cm$^2$, CNI Laser). Intensities were measured by LaserMate-Q (Coherent) intensity meter.

Reflectance spectra before and after irradiation were recorded using M266 (Solar Laser Systems, Belarus) spectrometer.

Refractive indices and birefringence of the photochromic copolymer **P3** were determined by polarized absorbance measurements of uniaxially aligned polyimide-coated electrooptical cell filled with mixture. Polarized absorbance spectra were measured using Unicam UV-500 UV-Vis spectrophotometer (with Glan-Taylor prism). Optical lengths were calculated using interference patterns in absorbance spectra [21] and extraordinary ($n_e$), ordinary ($n_o$) refractive indexes were determined.

**Acknowledgments**

This research was supported by the Russian Foundation for Basic Research and Czech Science Foundation according to the research project № 19-53-26007 (synthesis and study of the phase behaviour of the polymers, investigations of the photooptical properties of the composites), Czech Science Foundation [Project No. CSF 20-22615J], Ministry of Education, Youth and Sports of the Czech Republic [Project No. LTC19051] and Operational Programme Research, Development and Education financed by European Structural and Investment Funds and the Czech Ministry of Education, Youth and Sports [Project No. SOLID21 - CZ.02.1.01/0.0/0.0/16_019/0000760]. Authors also thank Dr. M. Bugakov for the molecular mass measurements and Dr. A. Stakhanov for **MAzo4** monomer synthesis.

**Conflict of Interest**

The authors declare no conflict of interest.

**Keywords**

Porous silicon, photonic crystal, liquid crystal, azobenzene, photoorientation

# Supporting information

# Photoinduced split of the cavity mode in photonic crystals based on porous silicon filled with photochromic azobenzene-containing substances

Alexey Bobrovsky, Sergey Svyakhovskiy, Valery Shibaev, Martin Cigl, Vera Hamplová, Alexej Bubnov

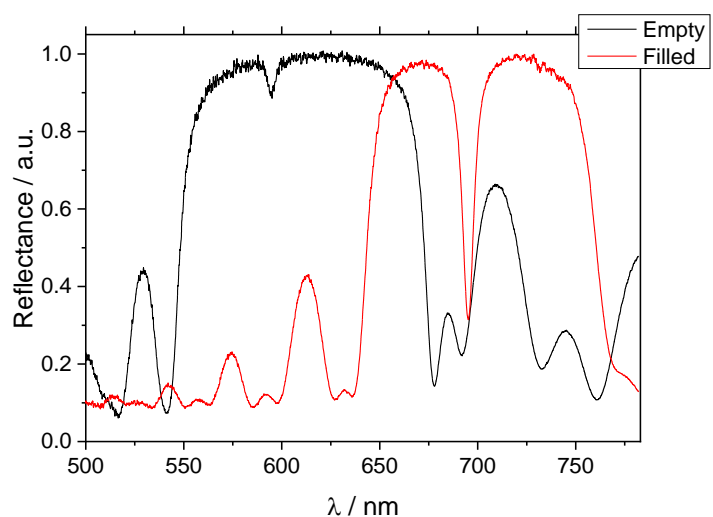

**Figure S1.** Reflectance spectra of the porous photonic structure before and after filling with polymer **P1**.

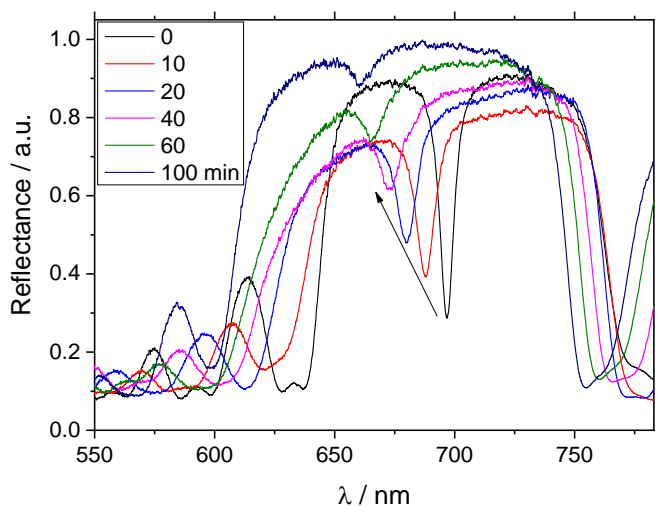

**Figure S2.** Reflectance spectra of the composite with **P1** during polarized UV light irradiation (355 nm, ~100 mW/cm$^2$).